\documentclass[twoside,fleqn,a4paper,11pt]{article}
\usepackage{epsfig}

\newcommand{\lsim}{\rlap{\raise 2pt \hbox{$<$}}{\lower 2pt \hbox{$\sim$}}}
\newcommand{\gsim}{\rlap{\raise 2pt \hbox{$>$}}{\lower 2pt \hbox{$\sim$}}}

\newcommand{\etal}{{\it et al.}}
\newcommand{\ic}{IC}
\newcommand{\ie}{{\it i.e.\ }}
\newcommand{\eg}{{\it e.g.\ }}
\renewcommand{\d}{\mbox{\rm d}}
\newcommand{\dbar}{{\bar{D}}}
\newcommand{\ccbar}{$c\bar{c}$}
\newcommand{\lc}{{\Lambda_c}}

\begin{document}
\noindent
TSL/ISV-96-0137      \hfill ISSN 0284-2769\\
April 1996           \hfill               \\
\\
\vspace*{5mm}
\begin{center}
  \begin{LARGE}
  \begin{bf}
Signals for Intrinsic Charm \\ in High Energy Interactions \\
  \end{bf}
  \end{LARGE}
  \vspace{5mm}
  \begin{Large}
G.~Ingelman$^{a,b,}$\footnote{ingelman@tsl.uu.se} and 
M. Thunman$^{a,}$\footnote{thunman@tsl.uu.se}\\
  \end{Large}
  \vspace{3mm}
$^a$ Dept. of Radiation Sciences, Uppsala University,
Box 535, S-751 21 Uppsala, Sweden\\
$^b$ Deutsches Elektronen-Synchrotron DESY,
Notkestrasse 85, D-22603 Hamburg, Germany\\
  \vspace{5mm}
\end{center}
\begin{quotation}
\noindent
{\bf Abstract:}
The prospects to test the hypothesis of intrinsic charm quarks in the
proton are investigated. We consider how this component can be directly
or indirectly probed in deep inelastic scattering at HERA and in fixed
target experiments and find that an overlooked signal might be present
in existing NMC data. Applying the intrinsic charm model to hadron collisions
we compare the resulting charm production cross-sections with those based on
standard perturbative QCD and available data. Extrapolating to higher energies 
we obtain predictions for charm production at the Tevatron and LHC.  
\end{quotation}

\section{Introduction}

The hypothesis of intrinsic charm quarks in the proton was introduced 
\cite{Brodsky80} as an attempt to understand a large discrepancy
between  early charm hadroproduction data and leading order
perturbative QCD (pQCD) calculations. The data from ISR (see refs.\ in
\cite{Brodsky80}) were about an order of magnitude higher than the
prediction and had a rather flat distribution in longitudinal momentum
compared to the sharp decrease with Feynman $x_F=p_{\parallel}/p_{max}$
expected from pQCD.  As more data have been collected at different
energies and next-to-leading  order (NLO) pQCD calculations have been
made, the discrepancy has largely disappeared. Still, however, there
are certain aspects of the data  which are difficult to understand
within the pQCD framework, but are  natural if the intrinsic charm
hypothesis is basically correct.  This concerns the mentioned $x_F$
distributions of leading charm in  hadroproduction
\cite{Vogt95b,Vogt95c}, the dependence on the nuclear number 
($A^{\alpha}$) of $J/\psi$ production \cite{Vogt91} and double $J/\psi$
production \cite{Vogt95a}. 

Thus, although the main features of charm production can  now be
understood by conventional pQCD, certain aspects of the data indicate
the presence of some additional mechanism. Intrinsic charm may be such a 
process, which gives a small contribution to the inclusive cross section 
but could dominate in some regions of phase space.   

The hypothesis of intrinsic charm (\ic) amounts to assuming the
existence of a \ccbar -pair as a non-perturbative component in the
bound state nucleon \cite{Brodsky80}. This means that the Fock-state
decomposition of the proton wave function,  $|p\rangle = \alpha
|uud\rangle + \beta |uudc\bar{c}\rangle +...$,  contains a small, but
finite, probability $\beta^2$ for such  an intrinsic quark-antiquark
pair. This should be viewed as a quantum fluctuation of the proton
state.  The normalization of the heavy quark Fock component is the key
unknown, although it should decrease as $1/m_Q^2$.  Originally, a 1\%
\ probability for intrinsic charm was assumed, but later 
investigations indicate a somewhat smaller but non-vanishing level; 
$\sim 0.3\%$ \cite{Hoffmann} and $(0.86\pm 0.60)\%$ \cite{Harris}. 

Viewed in an infinite momentum frame, all non-perturbative and thereby 
long-lived components must move with essentially the same velocity in 
order that the proton can `stay together'.  The
larger mass of the charmed quarks then implies that they take  a larger
fraction of the proton momentum. This can be quantified by applying
old-fashioned perturbation theory to obtain the momentum distribution
\cite{Brodsky80} 
\begin{equation} 
P(p\to uudc\bar{c}) \propto
\left[ m_p^2 - \sum_{i=1}^5 \frac{m^2_{\perp i}}{x_i} \right]^{-2}
\label{eq:IC-x}
\end{equation}
in terms of the fractional momenta $x_i$ of the five partons $i$  in
the $|uudc\bar{c}\rangle$ state.  The probability $\beta^2$ of this
state is  related to the normalisation $N_5$ of ref.~\cite{Brodsky80}. 
Neglecting the transverse masses ($m^2_{\perp}=m^2+p^2_{\perp}$) of the
light quarks in comparison to the charm quark mass results in the
momentum distribution 
\begin{equation}
P(x_1,x_2,x_3,x_c,x_{\overline{c}}) \propto 
\frac{x_c^2 x_{\overline{c}}^2}{(x_c + x_{\overline{c}})^2} 
\delta(1-x_1-x_2-x_3-x_c-x_{\overline{c}})
\label{eq:IC-simp}
\end{equation}
where the (transverse) charm quark mass is absorbed into the overall 
normalisation. This function
favours large charm quark momenta as anticipated. In fact, one
obtains $\langle x_c \rangle =2/7$ by integrating out the remaining
degrees of freedom $x_i$. 

An intrinsic \ccbar \ quantum fluctuation can be realised through an 
interaction, such that charmed particles are created. In
proton-proton collisions this could certainly happen through a hard
interaction with such a charm quark, but  the cross section is then
suppressed both by the small probability of the fluctuation itself and
by the smallness of the perturbative QCD cross section. The charm
quarks may, however, also be put on shell  through non-perturbative
interactions that are not as strongly suppressed \cite{Brodsky92}. Deep
inelastic lepton-proton scattering would be another possibility, where
this intrinsic charm component could be more directly probed.   To
investigate these possibilities we have constructed a  model based on
refs.~\cite{Brodsky80,Vogt91,Vogt92}. In part we use  Monte Carlo
techniques to simulate explicit events giving a powerful method  to
extract information on various differential distributions. 

One may also consider the extension from intrinsic charm to intrinsic 
bottom quarks. Since the overall probability scales with $1/m_Q^2$,
one expects all cross sections for intrinsic bottom to be about a factor 
ten lower. The $x$-distributions for bottom quarks should be somewhat harder
than for charm, but since light quark masses have already been neglected
in Eq.~(\ref{eq:IC-simp}) it also applies for intrinsic bottom and the same 
distribution as for charm is obtained in this approximation.     

In section 2 we investigate \ic \ in deep inelastic scattering (DIS)
in  fixed target and collider mode. In the former, we explicitly
consider the experiment New Muon Collaboration (NMC) at CERN and find
that an interesting number of intrinsic charm  events could be present
in their data samples. For  $ep$ collisions in HERA, we extend a
previous study \cite{IJN} to include the possibility of scattering on a
light quark such that the intrinsic charm \ccbar -pair  is freed and
give charmed particles in the forward proton remnant direction. The
case of hadronic charm  production is studied in section 3. The
measured charm production cross  sections are compared to those
calculated in pQCD to constrain  the allowed amount of IC. Here, an
important issue is the energy  dependence of the cross section for IC,
where two alternatives are considered.  Based on this, we provide
differential distributions of charm production from  \ic\ in comparison
to the standard pQCD treatment, as applied to  the Tevatron and LHC.
Finally, section 4 gives a concluding discussion. 

\section{Intrinsic charm in deep inelastic scattering}
\subsection{General framework}

\begin{figure}
\begin{center}
\includegraphics[width=13cm]{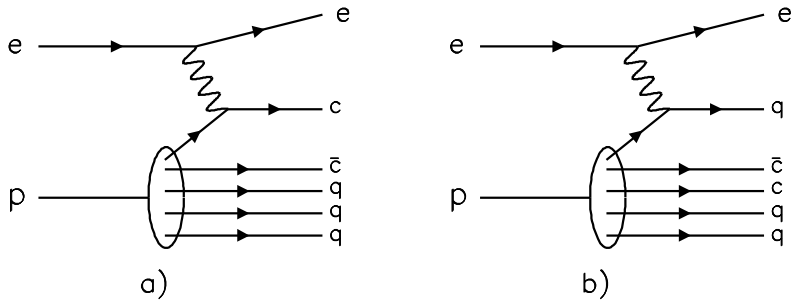}
\caption[junk]{\it 
Deep inelastic scattering on a proton containing an intrinsic
$c\bar{c}$-pair; a) scattering on the charm quark, b)
scattering on a light valence quark.}
\label{fig:Feynman}
\end{center}
\end{figure}

In DIS it should be possible to measure the effective \ic \
parton density by direct scattering on an
intrinsic (anti)charm quark (Fig.\,\ref{fig:Feynman}a).
The parton density is obtained from Eq.\,(\ref{eq:IC-simp})
by integrating out all degrees of freedom except the momentum fraction $x$ 
of the charm quark (or antiquark), resulting in 
\begin{equation}\label{eq:c(x)}
c_{IC}(x)=\beta^2 1800x^2\left\{
\frac{1}{3}(1-x)(1+10x+x^2)+2x(1+x)\ln{x}\, \right\}
\end{equation}
The DIS charm cross section is given by 
\begin{equation}\label{eq:DIS-sigma}
\frac{\d ^2\sigma}{\d x \d Q^2} = \frac{2\pi \alpha}{xQ^4}
\left[1+(1-y)^2\right]\, F_2^c(x,Q^2)
\end{equation}
in terms of the conventional DIS variables; Bjorken $x=Q^2/2P\cdot q$, 
$y=P\cdot q/P\cdot p_\ell$ where $P,p_\ell$ is the momentum of the initial 
proton and lepton, and $q$ is the four momentum transfer of the exchanged 
photon. Only photon exchange contribution is here included since the 
high-$Q^2$ region with $Z^0$ exchange contributions can be neglected in 
our applications.  The general charm structure function is  
\begin{equation}\label{eq:F2c}
F_2^c(x,Q^2) = e_c^2\, \left\{ xc(x,Q^2)+x\bar{c}(x,Q^2)\right\}
\end{equation}
and $F_2^{IC}$ is obtained if the intrinsic charm quark
density $c_{IC}(x,Q^2)$ is used. The $Q^2$ dependence from 
normal leading log GLAP equations have been calculated for IC in
\cite{Hoffmann}, but can be taken into account through a  simple extension 
of the parameterisation in Eq.~(\ref{eq:c(x)}) \cite{IJN}.

We have included an option to treat  DIS on intrinsic heavy quarks  in
the Monte Carlo program {\sc Lepto} \cite{Lepto} using this formalism
and are therefore able to simulate complete events based on this \ic \
model. As for normal DIS events, higher order perturbative parton emissions 
from the incoming and scattered quark are included through parton showers. 
Hadronisation and particle decays are then performed
according to the  Lund model \cite{Lund} in its Monte Carlo
implementation {\sc Jetset} \cite{Jetset}.

In addition to this direct scattering on an intrinsic charm quark,  one
may also consider the case of DIS on a light (valence) quark
(Fig.\,\ref{fig:Feynman}b) such that 
the intrinsic \ccbar -pair in the proton is `liberated' and give
charmed  particles in the phase space region corresponding to the
`spectator' proton remnant. To simulate this case also, we have made
add-on routines to {\sc Lepto} based on the following simple model.  The
normal electroweak DIS cross section is first used to treat the basic 
scattering on a light quark. The corresponding quark density function
is  here obtained from Eq.\,(\ref{eq:IC-simp}) by integrating out all
degrees of freedom,  except $x_1$ for the light (valence) quark to
scatter on,
\begin{equation}
q_{IC}(x_1)=\beta^2\, 6\,(1-x_1)^5.
\end{equation}
The probability $\beta^2$ of the $|uudc\bar{c}\rangle$ state is included here 
and gives the normalisation of the cross section for this intrinsic charm
process.   

\begin{figure}
\begin{center}
\includegraphics[width=13cm]{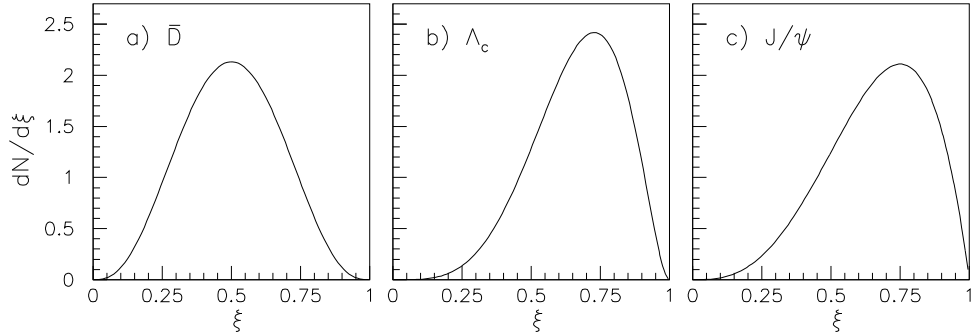}
\caption[junk]{\it 
Distribution of rescaled longitudinal momentum $\xi=x/(1-x_1)$ of  (a)
$\bar{D}$-mesons, (b) $\Lambda_c$ and (c) $J/\psi$ produced through 
coalescence in the remnant system from a DIS scattering on a valence
quark  in the intrinsic charm $|uudc\bar{c}\rangle$ state. The distributions
shown in the scaled variable are normalised to unit integral as discussed in 
the text.}
\label{fig:disIC}
\end{center}
\end{figure}

From the remaining $qqc\bar{c}$ system, one then considers the
production of $\bar{D}$-mesons, $\Lambda_c$ and $J/\psi$ through a
coalescence \cite{Vogt91,Vogt92} of the relevant partons 
in Fig.\,\ref{fig:Feynman}b. The
longitudinal momentum distributions for these charmed particles are
obtained, for a given $x_1$, by integrating over the other
momentum fractions in Eq.\,(\ref{eq:IC-simp}) and inserting a 
$\delta$-function for the particle to be produced, \ie
$\delta(x_{\bar{D}}-x_q-x_{\bar{c}})$,
$\delta(x_{\Lambda_c}-x_u-x_d-x_c)$  or
$\delta(x_{J/\psi}-x_c-x_{\bar{c}})$.  Considering momentum fractions
relative to the proton remnant,  \ie after $x_1$ is removed,
corresponds to the variable substitution  $x\to \xi=x/(1-x_1)$. This
gives expressions that  factorise into an $x_1$-dependent part and a
$\xi$-dependent part.  Since $x_1$ is already chosen, the former part can be
absorbed into the  normalisation of the $\xi$-distribution.
Thus, normalising to overall unit probability, one 
obtains the probability distributions
\begin{eqnarray}
\frac{\d N_{\dbar}}{\d \xi} & = & P_{\dbar}\ 300\ \left\{\
\xi^4\,\ln\, \xi\ +\ (\,1-\xi\,)^4\,
\ln\,(1-\xi)\ + \begin{array}{c} \ \\ \ \end{array} \right. 
\nonumber \\ 
& & \left. \xi \,
(\,1-\xi\,)\, [\xi^2-\frac{1}{2}\xi 
(\,1-\xi\,)+(\,1-\xi\,)^2]\  
\begin{array}{c} \ \\ \ \end{array} \right\}
\\
\frac{\d N_{\lc}}{\d \xi} & = & P_{\lc}\ 300\ (\,1-\xi\,)^2\
\left\{ \xi\,[\,6\, -5\xi\,]\ +
2\,[\,\xi^2\,+\,4\xi\,+\,3\,]\ 
\ln (1-\xi) \right\}
\\
\frac{\d N_{J/\psi}}{\d \xi} & = & P_{J/\psi}\ 20\ (\,1-\xi\,)
\,\xi^3
\end{eqnarray}
where $P_{\dbar}, P_{\lc}$ and $P_{J/\psi}$ are the relative
probabilities  to form the three different charmed particles.  These
distributions are illustrated in Fig.\,\ref{fig:disIC}. 

The overall normalisation is already accounted for (by $\beta^2$) in
the DIS cross section and only the relative probabilities $P_{\dbar}$,
$P_{\lc}$ and $P_{J/\psi}$ need to be specified. Following the
investigations in \cite{Vogt91,Vogt92} for the similar case in
hadroproduction, we take 66\% $\overline{D}$, 33\% $\Lambda_c/\Sigma_c$
and 1\% $J/\psi$. In each case there is a remaining parton in the
proton remnant, \ie a $qc$ diquark, a $\bar{c}$ quark and a $qq$
valence diquark, respectively. These are all anti-triplets  in colour
and are connected with a Lund string to the scattered quark to form a
singlet system that hadronises using the normal Lund model. Note, that
the charm (anti)quark at the proton remnant end of the string also
gives rise to a charmed hadron. Before hadronisation, however, the
scattered quark may radiate partons as treated by parton showers
in {\sc Lepto}.  

In the following, we extend the previous study \cite{IJN} of  direct
DIS on \ic \ in two ways.  First, a more dedicated study of direct \ic
\ scattering in  connection with the New Muon Collaboration (NMC)
experiment is made.  Secondly, we consider the above model for indirect
\ic\ scattering applied to  $ep$ collisions at HERA and compare with
direct \ic \ scattering as  well as conventional boson-gluon fusion
into \ccbar \ from pQCD.   

\subsection{Fixed target muon scattering (NMC)}

In case of fixed target experiments the interest is focused on the
direct scattering on intrinsic charm. This gives a high energy charmed
particle in the current direction. The resulting high energy muon from
a semi-leptonic decay can then be used as a signal for
charm. In the indirect \ic \ process, however, the charmed particles
will emerge in the target region and resulting muons will have low
energy and hence be more difficult to identify. Therefore we will not
consider this indirect  mechanism in the fixed target case.

We thus concentrate on the direct scattering on intrinsic charm quarks 
and use our Monte Carlo model to simulate events corresponding to the 
NMC experimental situation. This means having a $280\: GeV$ muon beam
on  a stationary proton target and applying the cuts  
$Q^2>2\,GeV^2$ and $W^2>100\,GeV^2$.

\begin{figure}
\begin{center}
\includegraphics[width=6cm]{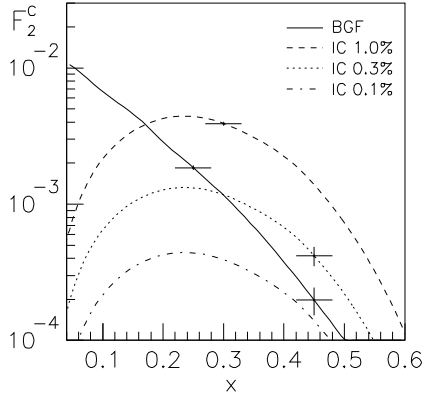}
\caption[junk]{\it 
The charm structure function $F^c_2(x)$ obtained from our Monte Carlo 
simulation of DIS on intrinsic charm quarks, with the different 
indicated normalisations ($\beta^2$), and the conventional pQCD boson
gluon fusion process  $\gamma^{\ast} g\to c\bar{c}$ (solid curve). The
error bars indicate the potential statistical precision of the
discussed NMC data sample.}
\label{fig:nmcf2}
\end{center}
\end{figure}

From the simulated intrinsic charm events we extract the structure
function $F_2^{IC}$, using Eq.\,(\ref{eq:DIS-sigma}), resulting in the
three curves in Fig.~\ref{fig:nmcf2} for different intrinsic charm 
probabilities. For comparison, we have calculated $F_2^c$ arising from
the conventional boson-gluon fusion (BGF) process  $\gamma^{\ast} g\to
c\bar{c}$. This was obtained from a similar Monte Carlo simulation
using the {\sc Aroma} program \cite{Aroma}, which is an  implementation
of this process using the leading order pQCD matrix elements with the
proper charm quark mass effects included (together with parton showers
and hadronisation as in {\sc Lepto}). The large-$x$ feature of \ic \ is
clearly seen and results in a dominance of \ic \ over pQCD at large
enough $x$. The cross-over point depends on the  unknown absolute
normalisation of \ic \ as illustrated. 

One should here consider the experimental information on the inclusive 
$F_2^c$ available from the European Muon Collaboration (EMC). 
Their original analysis  gave an
upper limit for IC ($\beta^2$) of 0.6\% at 95\% CL \cite{Aubert82}. In
a later analysis \cite{Hoffmann}, the QCD evolution of the \ic \ 
structure function, as mentioned above, and charm mass effects were  
taken into account giving positive evidence for \ic \ at the level
0.3\%.  A very recent investigation \cite{Harris} (in parallel with our
study)  has improved the theoretical treatment by including
next-to-leading order  corrections both for the \ic \ and pQCD
processes. Based on the same  EMC data they find that an intrinsic
charm contribution of  $(0.86\pm 0.60)\%$  is indicated (in the bin of
large energy transfer with mean $\nu =168\,GeV$).

Thus, there is some evidence for \ic \ from the inclusive $F_2^c$
measurement  of EMC, but it is not conclusive. One therefore needs
to consider whether further information can be obtained. 

The success of the experimental method to tag charm through muons
depends  very much on the amount of muons from other sources.  The
background to \ic \ from perturbatively produced charm, as given above,
is in a sense an irreducible background. There is, however, also a
severe  background from the decays of light hadrons, mostly pions and
kaons, that are copiously produced. Since these particles have a much
longer life-time than  charm, their effect can be reduced by either
having the muon detectors close to the target or having hadron
absorbers to filter out the prompt muons. 

The normal experimental arrangement of NMC is to have little material 
between the target and the detectors. This
implies, however, a substantial muon background from $\pi$ and $K$ decays.  
However, in some runs of the NMC experiment a set-up more suitable for
a charm search was used. In the heavy target configurations during 
1987-88 there were hadron absorbers immediately downstream of
the targets. We have developed an analytic method to
calculate the suppression of this muon background due to the hadron
absorption. The details are reported in \cite{Thunman96}, where it is
also applied to this NMC set-up. Thus, we calculate the
development of the flux of produced mesons within the targets, in the
calorimeters/absorbers and in the empty space between these elements.
The resulting muon flux from hadron decays was 
found to be reduced to only $\sim6\%$ of the flux without such
absorption. In addition, we found that the calorimeter between the
target and the absorber could also be used as a target. Charmed hadrons
can also be absorbed, but due to their much shorter life-time the
effect is negligible. For example, the decay length of a $D^{\pm}$ is 
at most $\sim 1/10$ of its interaction length (corresponding to a momentum
of $180\,GeV$, which is about the highest to be expected). 

\begin{figure}
\begin{center}
\includegraphics[width=14cm]{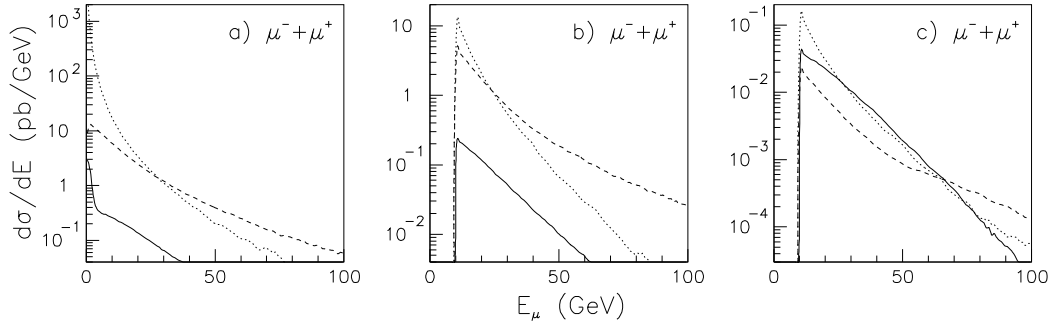}
\caption[junk]{\it 
Differential cross section versus energy of muons ($\mu^-\,+\,
\mu^+$) from the decay of charmed particles originating from a 1\% 
intrinsic charm component in the proton (solid curves), 
from pQCD $\gamma^{\ast} g\to c\bar{c}$
(dashed curve) and from  decay of light mesons (mainly $\pi , K$)
(dotted curves); all in  $280\: GeV$ muon-proton scattering with cuts:
(a) $Q^2>2\,GeV$ and $W^2>100\,GeV^2$,  
(b) NMC cuts on the scattered muon $42<E_{\mu 1}<224\: GeV$ and  
$\theta > 10\,mrad$ and on secondary muons $E_{\mu 2}>10\,GeV$,  
(c) $E_{\mu 1}>100\,GeV$ and $x_{Bj}>0.2$ to enhance IC.
An overall reduction for muons from $\pi$ and $K$ decays to 6\%
has been applied to account for the discussed hadron absorption effect
in NMC.}
\label{fig:emcspec}
\end{center}
\end{figure}

An experimentally realistic study of signal and background muons  has
been performed using Monte Carlo simulations. The results are displayed
in  Fig.~\ref{fig:emcspec}.  The background from light hadron decays
was obtained by applying the above  suppression factor on the yield
from simulated  normal DIS events.  Here, the standard version of {\sc
Lepto} was used  in the same kinematic region as specified above.  The
muons from \ic \ and the pQCD process are obtained with our modified 
{\sc Lepto} version and {\sc Aroma}, respectively. 

The resulting energy spectra of muons are shown in
Fig.\,\ref{fig:emcspec} for three sets of cuts. Fig.\,\ref{fig:emcspec}a
is for the above mentioned standard cuts; $Q^2>2\,GeV$ to ensure
DIS and $W^2>100\,GeV^2$ to have enough energy in the hadronic system
for charm pair production. Fig.\,\ref{fig:emcspec}b have extra cuts
motivated by the NMC experiment; the scattered muon in the energy range
$42<E_{\mu 1}<224\: GeV$ and at angle $\theta > 10\,mrad$ to emerge
from the beam and secondary muons (from other sources) having a large
enough energy $E_{\mu 2}>10\,GeV$ to be indentified. In Fig.\,\ref{fig:emcspec}c
the additional cuts $E_{\mu 1}>100\,GeV$ and $x_{Bj}>0.2$
were also made to identify the highest energy muon as the scattered one
and select the large-$x$ region. This last requirement is ment to
enhance intrinsic charm events over the background processes ({\em cf.}
Fig.\,\ref{fig:nmcf2}). As demonstrated, it is then possible get muons
from IC to the same level as the backgrounds. 

With the additional cuts the absolute rate is lowered such that large 
statistics data samples are needed. This may be satisfied by the NMC
run in 1988 when data were collected for $9.5\cdot10^{11}$ muons of
$280\,GeV$ on four targets of thickness $145\,g/cm^2$ each, which
corresponds to an integrated luminousity of $334\,pb^{-1}$. The
statistical precision that can be expected in $F_2^c$ for this sample
is illustrated in Fig.\,\ref{fig:nmcf2} for a few $x$-bins. This
should give $\sim 200$ events with muons ($\mu^-+\mu^+$) of energy
larger than $10\,GeV$ and fulfilling the stricter set of cuts, \ie
corresponding to the integrated distribution in
Fig.\,\ref{fig:emcspec}c. In the same way one obtains $\sim 85$
muons from charm produced by the pQCD process and $\sim 340$ muons from
$\pi ,K$ decays after the hadron absorpion reduction as estimated with
the above mentioned analytic method. Additional data based on
$8.4\cdot10^{11}$ muons of $200\,GeV$ energy could also be analysed in
this way.

In order to reduce the background from $\pi , K$ decays further, one
might also use secondary vertices based on 
the difference in decay length of light mesons as compared to
charmed  particles. We have investigated \cite{Thunman96} this by
considering the  impact parameter of the muon tracks with respect to
the primary interaction  vertex. The impact parameter for muons from
charm is always less than  $0.5\: cm$, whereas it is typically larger
for those from light mesons.  However, the hadron absorbers that
reduce the rate of muons from light mesons, also give rise to multiple
scattering of the muons passing through. This causes a smearing of the
reconstructed impact parameter which we find \cite{Thunman96} has a
Gaussian distribution of width $3.5\: cm$. Thus, the {\em measureable} 
impact parameter of muons from charm does not seem to provide a useable 
signature, although a safe conclusion would require a detailed simulation 
of the experimental conditions. 

In these estimates for IC we have assumed a 1\% normalisation
($\beta^2$) of the IC component and leave a trivial rescaling to  any
smaller value that may be preferred. Based on the above, we conclude that 
it seems possible to find an intrinsic charm signal down to the 0.2\% level
in the NMC data through an excess of muons as compared to the expected
background (\eg  an excess of about 40 over a background of 425 muons). 
The same sensitivity can be estimated from the precisions of $F_2^c$ 
as shown in Fig.\,\ref{fig:nmcf2}. The determination of the BGF background
is here essential, but uncertainties in the overall normalisation of the 
theoretical calculation can be well determined at 
small-$x$ where this process dominates $F_2^c$ and the statistical precision 
is high. 

One should note that at large $Q^2$, the intrinsic and extrinsic (BGF)
charm processes have different scaling behaviour with the charm quark mass; 
$1/m_c^2$ and $log(m_c^2)$, respectively. This can give an additional handle 
to separate them in a data sample. 

\subsection{Electron-proton collider}

In HERA at DESY, $30\,GeV$ electrons (or positrons) are collided with 
$820\; GeV$ protons. The direct DIS on intrinsic charm quarks in the
proton  has been investigated previously \cite{IJN}. 
Since the intrinsic charm quark typically has a  large
$x$, it will be scattered at a rather small forward angle close to
the  proton beam direction. This makes it difficult to detect muons
from  semileptonic decays and only a fraction of the events are 
observable. For details we refer to \cite{IJN}. 

\begin{figure}[b]
\begin{center}
\includegraphics[width=15cm]{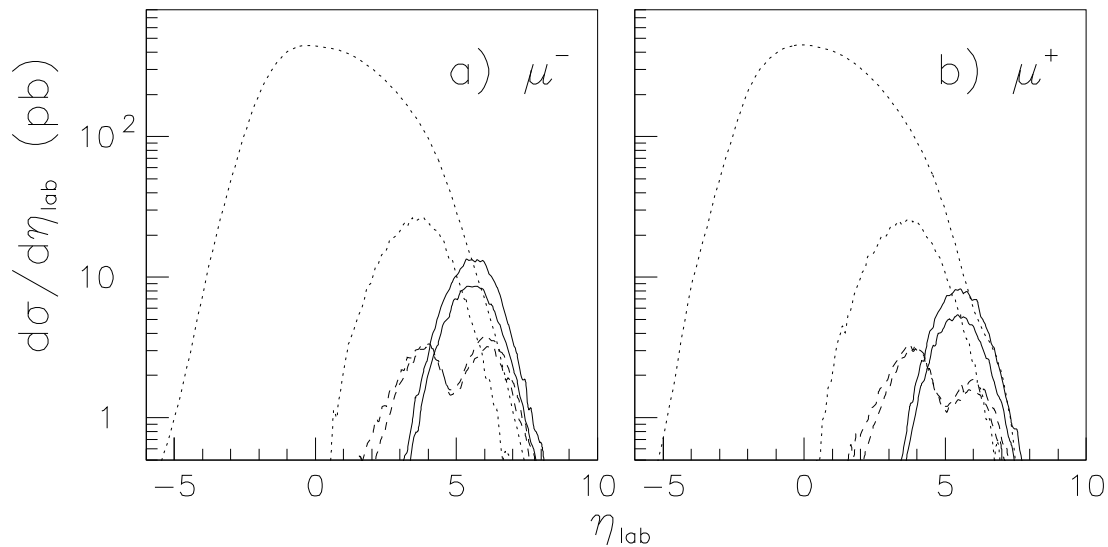}
\caption[junk]{\it
Distribution in pseudorapidity ($\eta = - ln\ tan(\theta /2)$) for
muons from  semileptonic decays of charmed particles produced through
different DIS  processes at HERA: scattering on a light (valence) quark
in the intrinsic  charm proton state $|uudc\bar{c}\rangle$ (solid
curves), direct scattering on  an intrinsic charm state (dashed curves)
and the conventional pQCD   $\gamma^{\ast} g\to c\bar{c}$ process
(dotted curves). In each case, the upper  curve correspond to the
inclusive event sample (see text) and the lower  curve after the cuts
$x>0.03,\: y<0.3$ made to suppress the pQCD contribution.
(Note, $\eta>0$ corresponds to the proton beam direction.)}
\label{fig:heraic}
\end{center}
\end{figure}

The above discussed case of indirect intrinsic charm scattering may
also  occur. The charm particles are then produced in the
forward-moving proton  remnant system and will emerge in the very
forward direction and therefore be hard to detect. Still, it is worthwile to
estimate the rate and the distribution in pseudo-rapidity of the
emerging muons.

In a recent paper \cite{BTH} these different IC processes in $ep$ collisions
were discussed at a theoretical and qualitative level. Here, we concentrate
on a phenomenological study of these processes in order to get quantitative
results for direct experimental considerations.    

It is of obvious interest to compare the different charm production 
mechanisms, which can easily be made through our different Monte Carlo 
simulation programs as discussed above. Thus, we use the add-on to {\sc
Lepto} to simulate the indirect IC scattering process,  {\sc Lepto} for
direct scattering on an intrinsic charm quark and  {\sc Aroma} for the
pQCD $\gamma^{\ast} g\to c\bar{c}$ process.  In all cases $ep$ events
at the HERA energy were simulated with $Q^2>4\,GeV^2$ and 
$W^2>100\,GeV^2$.

In Fig.\,\ref{fig:heraic} we display the resulting distributions in 
pseudo-rapidity for muons from charm
decays in  the different cases.   As can be seen, the conventional
pQCD process dominates strongly the overall rate, but can be
substantially reduced by the simple cuts  $x>0.03,\: y<0.3$ which
hardly affects the IC contributions.  The two IC processes 
give charm at large  rapidities in the proton beam direction, as
expected. 

One may notice that  the muon detectors in the HERA experiments only
covers pseudo-rapidities up to $3-4$. 
A possibility would be to lower the proton beam energy and thereby decrease
the strong forward boost effect. For example, lowering the proton beam 
energy to $300\; GeV$, would essentially shift the distributions lower 
in rapidity by about one unit. This has been investigated for the direct 
IC scattering in ref.~\cite{IJN}, where also the background of muons from 
light meson decays ($\pi ,K$) was studied. 

Although it does not seems possible to detect these IC processes with todays
set-ups in the H1 and ZEUS experiments, it is of interest for the
upgrades under discussion, \eg within the presently running workshop
on `Future Physics at HERA'. 

\section{Intrinsic charm in hadron interactions} 

We now turn to hadron collisions for which the intrinsic charm
hypothesis  was first developed \cite{Brodsky80}. Here also, one could
consider to probe the intrinsic charm quarks in hard scattering
processes such that pQCD is applicable. This would, however, result in
a very low cross section; suppressed both by the IC probability and the
smallness of the perturbative cross section.  The charm quark may,
however, also be put on shell and emerge in real charmed particles
through softer non-perturbative interactions that are not strongly
suppressed \cite{Brodsky92}. As in previous investigations of this subject
\cite{Vogt91,Vogt92},  we will consider the shapes of $x_F$
distributions as derived from the  IC model separately from 
the overall  normalisation of the cross section. The energy dependence
of the cross section is a non-trivial issue, which we discuss in connection
with pQCD charm production and charm hadroproduction data.

\subsection{Basic $x_F$ distributions in IC}

A non-perturbative hadronic interaction of a proton in an IC state  is
thus assumed to realise charmed hadrons.  This may occur through
conventional hadronisation in a similar way as  when the charm quark
and antiquark is produced in a hard interaction,  \eg in a
hadron-hadron or $e^+e^-$ collision. This can be well describe  by a
fragmentation model, like the Lund model \cite{Lund}, or a
parametrised fragmentation function such as the Peterson function
\cite{Peterson}.  The effective charm fragmentation function is quite
hard such that a simple  approximation is to use a $\delta$-function
to let the charmed hadron get the same momentum as the charmed quark.
Given the uncertainties of the intrinsic  charm momentum distribution
to be used as input for the fragmentation,  this $\delta$-function
fragmentation is adequate here and has also been used in previous IC
studies \cite{Vogt95c}. 

The charm (anti)quark may also coalesce with another quark or diquark
from the same initial proton state $|uudc\bar{c}\rangle$ to form a
hadron. This may happen when such a parton system forms a colour
singlet, but has too small invariant mass to hadronise according to the
Lund model. Such a mechanism is especially important for the production
of $\Lambda_c$, but should also play an important role in the
production of $D$-mesons. For an initial proton, this gives an
asymmetry between $\overline{D}$ mesons, which can be formed by this
coalescence process, and $D$ mesons which cannot. Such differences in
leading particle spectra are observed experimentally (see
\cite{Vogt95c} and references therein) and may be an indication
for these mechanisms. The $c$ and $\bar{c}$ can also coalesce to form a
charmonium state, mainly $J/\psi$ or some higher resonance that mostly
decays into the lowest lying $J/\psi$ state. 

The relative probabilities for these processes are not known, but have
been discussed before. Following \cite{Vogt91,Vogt92} we use the
recombination probabilities 50\% \ to form a $\overline{D}$-meson and
30\% \ for a $\Lambda_c$. The probability to directly form a $J/\psi$
(\ie the \ccbar \ pair is combined) is taken to be 1\%. The absolute
rates of $\Lambda_c$ and $J/\psi$ do, of course, depend strongly on
these values.  The $D$-meson rates and distributions are, however, not
so sensitive to these values because the resulting distributions from
fragmentation and coalescence are rather similar. The $c$ or $\bar{c}$
quarks that do not coalesce with spectator partons, are hadronised to 
$D$-mesons with the mentioned $\delta$-function. 
This is the case for the remaining $c$ or $\bar{c}$-quark in events with 
coalescence and both $c$ and $\bar{c}$ in the remaining 19\% of events 
without coalescence.

The momentum of the hadron formed through coalescence is taken as the
sum of the corresponding $x_i$'s, \eg $x_{\Lambda_c}=x_c+x_u+x_d$. The
momentum distribution is then obtained by folding
Eq.\,(\ref{eq:IC-simp}) with the proper $\delta$ function, \eg 
$\delta(x_{\Lambda_c}-x_c-x_u-x_d)$, and integrating out all extra
degrees of freedom. When the $c$ or $\bar{c}$ quarks are hadronised 
with the $\delta$-function, the $D$-meson takes the whole intrinsic
charm quark momentum as given by Eq.\,(\ref{eq:c(x)}). 
This procedure is consistent with low-$p_t$ charm hadroproduction data
\cite{Vogt95c}.

\begin{figure}[t]
\begin{center}
\includegraphics[width=12cm]{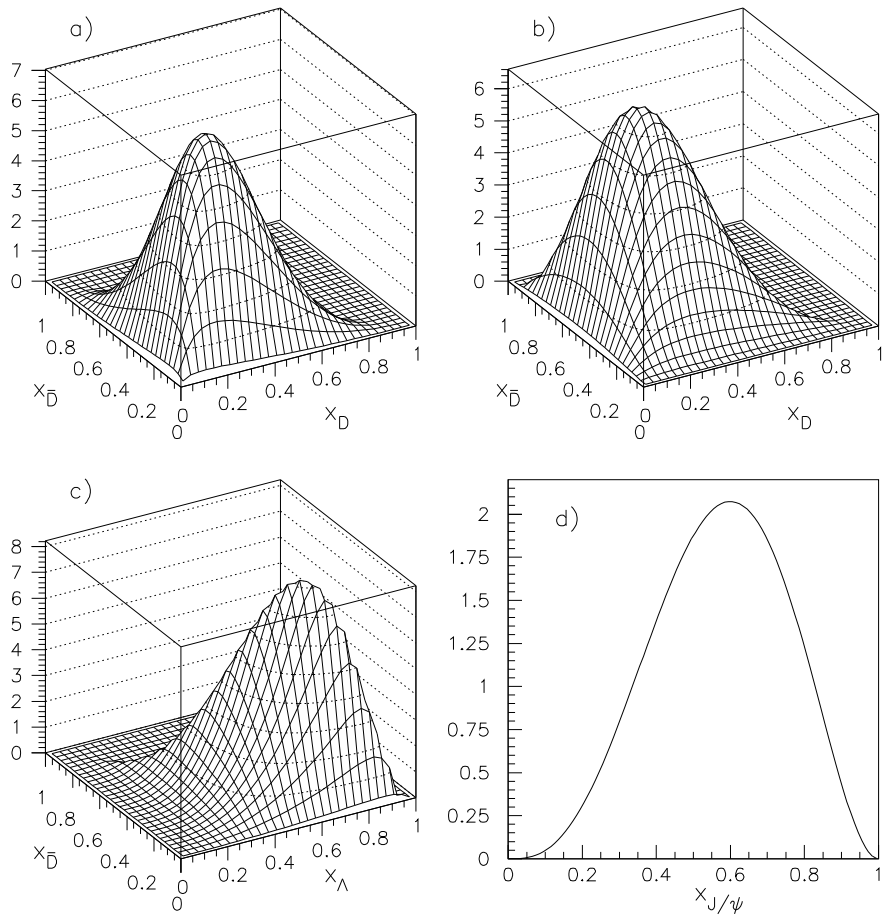}
\caption{\it 
Probability distributions $dP/dx_1d\,x_2$ and $dP/dx_{J/\psi}$ in
longitudinal momentum fractions  ($x_i$) for  charmed particles from
IC:  (a) $D\, \overline{D}$ from fragmentation,
Eq.\,(\ref{eq:DDfrag}).  (b) $D$ from fragmentation and a
$\overline{D}$ from coalescence,  Eq.\,(\ref{eq:DD}). (c)
$\overline{D}$ from fragmentation and $\Lambda_c$ from coalescence, 
Eq.\,(\ref{eq:DLambda}). (d) $J/\psi$ from coalescence,
Eq.\,(\ref{eq:psi}).}
\label{fig:dndxdx}
\end{center}
\end{figure}

Based on this model we then obtain the following correlated
probability  distributions in longitudinal momentum fraction
(Feynman-$x$) for the  different combinations of charmed hadrons in an
event; a $D\, \overline{D}$ pair produced through fragmentation      
\begin{equation}\label{eq:DDfrag}
\frac{\d P}{\d x_D\,\d x_{\bar{D}}}  \propto 
1800\ \frac{(1-x_D-x_{\bar{D}})^2 x_D^2 x_{\bar{D}}^2}
{(x_D+x_{\bar{D}})^2}\: ;
\end{equation}
a $D$ from fragmentation and a $\overline{D}$ from coalescence
\begin{equation}\label{eq:DD}
\frac{\d P}{\d x_D\,\d x_{\bar{D}} }  \propto 
3600\ (1-x_D-x_{\bar{D}}) x_D^2
\left(x_D+x_{\bar{D}}-\frac{x_D^2}{x_D+x_{\bar{D}}}- 2 x_D^2 \ln
\frac{x_D+x_{\bar{D}}}{x_D^2} \right)\: ;
\end{equation}
a $\overline{D}$ from fragmentation and a $\Lambda_c$ from coalescence
\begin{equation}\label{eq:DLambda}
\frac{\d P}{\d x_{\Lambda_c}\,\d x_{\bar{D}}}  \propto 
3600\,x_{\bar{D}}^2 \left( \frac{x_{\Lambda_c}}{2} +
3\,x_{\bar{D}}\,x_{\Lambda_c} -  x_{\bar{D}}\,(3\,x_{\bar{D}} +
2\,x_{\Lambda_c})\,\ln\frac{x_{\bar{D}}+x_{\Lambda_c}}{x_{\bar{D}}}
\right)\: ;
\end{equation}
and a $J/\Psi$ from coalescence
\begin{equation}\label{eq:psi}
\frac{\d P}{\d x_{J/\psi}}  \propto  
20\,(1-x_{J/\psi})^2 x_{J/\psi}^3.
\end{equation}
These two-dimensional distributions are plotted in Fig.\,\ref{fig:dndxdx}. 
Each one is here normalised to unit integral and their relative weight
are the percentages discussed. Obviously, the sum of
the two $x$'s in each case cannot exceed unity and the distributions
therefore vanish when  passing the diagonal in the two-dimensional
plots. The $J/\psi$, formed by the coalescence of the $c\bar{c}$-pair, 
take a larger momentum fraction since it is formed of two quarks with
relatively high $x_F$, as shown in Fig.\,\ref{fig:dndxdx}. The
$\Lambda_c$, which coalesce with two valence quarks, gets a  larger momentum
than the $D$ mesons as demonstrated in Fig.\,\ref{fig:dndxdx}.

In addition to these charmed particles, higher mass states may also
be  produced. However, these decay rapidly to the treated ones and can
therefore be  considered included in these
parameterisations. $D_s$ cannot be produced in the coalescence
process, but is allowed in the fragmentation  process although at a
suppressed rate (\eg by a factor $\sim 7$ in the Lund model). Since
its spectrum would be essentially the same as the other  $D$-mesons,
we include it with them and use $D$ as a generic notation for  all
pseudoscalar charm mesons.   The vector mesons $D^*$ decays strongly
to $D$-mesons, but the decays are  not charge symmetric; $D^{*0}\to
D^0 (100\%),  D^{*\pm}\to D^0 (68\%)/D^{\pm}(32\%)$.  
This effect is taken into account after having chosen a primary $D^*$ or
$D$ from spin statistics.  

The shapes of the $x_F$-distributions for charmed particles from
intrinsic charm are thereby specified and found to be quite hard, as
expected. In fact, they are harder than those for charm from
perturbatively produced charm, as shown below. This gives a possibility
that IC may contribute substantially or even dominate at large $x_F$,
depending on the absolute normalisation of the production cross
section. 

\subsection{Normalisation and energy dependence of the IC cross section}

\begin{figure}[t]
\begin{center}
\includegraphics[width=16cm]{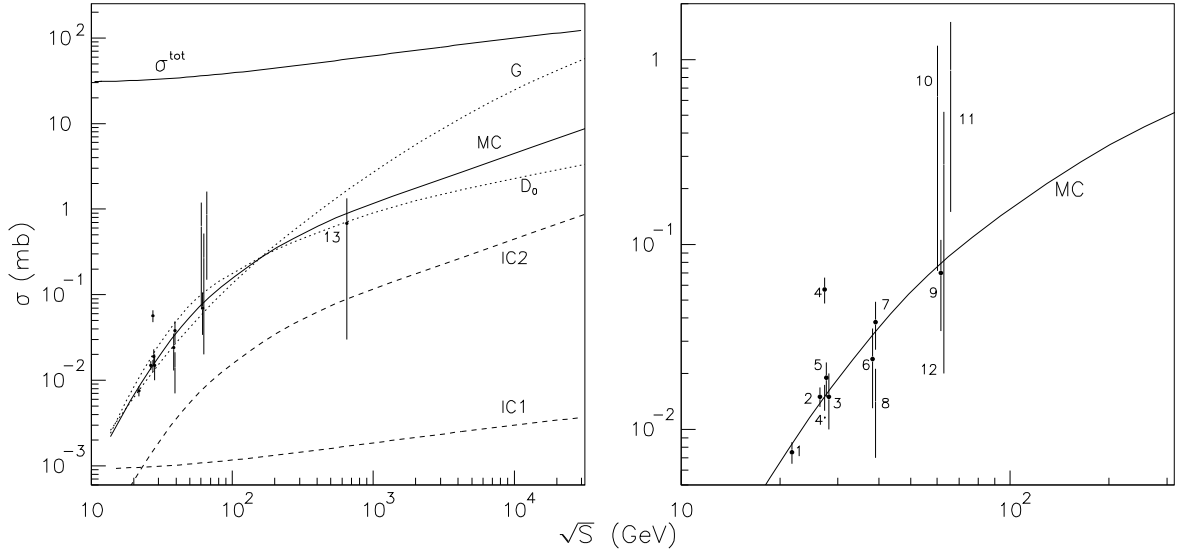}
\caption[junk]{{\it Energy dependence of charm production cross section in $pp
(p\bar{p})$. The experimental data points are 1:\cite{E769a}
2:\cite{lebc400}, 3:\cite{Jonker}, 4,4':\cite{Duffy}, 5:\cite{Fritze}, 
6:\cite{lebc800}, 7:\cite{E653}, 8:\cite{E789} 9:\cite{Clark},
10:\cite{r608}, 11:\cite{Basile1,Basile3},  12:\cite{Basile2,Basile3}
and 13:\cite{ua2}.  The solid line (MC) is the result of our Monte
Carlo calculations based on  conventional pQCD and hadronisation, with
dotted lines from variations   using a naive application of the $MRS$
parametrisations $G$ and $D_0$  of  parton densities in the proton. 
The dashed curves are from our model for intrinsic charm, with the two 
assumptions IC1 and IC2 about its energy dependence. For reference,
the total $pp$ cross section $\sigma^{tot}$ is also shown.}}
\label{fig:cross}
\end{center}
\end{figure}

As mentioned, the main uncertainty in the intrinsic charm model is the
absolute normalization of the cross section and its energy dependence. 
Since the process is basically a soft non-perturbative one, it may
be reasonable to assume that its energy dependence is the same as for
normal inelastic scattering \cite{Brodsky92}. We therefore take as our
first case
\begin{equation}
{\rm IC1:}\;\;\; \sigma_{IC}(s) =  3\cdot 10^{-5} \sigma_{inel}(s)
\label{eq:IC1}
\end{equation}
shown as curve IC1 in Fig.\,\ref{fig:cross} and with  normalisation
from ref.~\cite{Vogt92}, where the magnitude of the cross section  was
estimated from data at relatively low energies ($\sqrt{s}=
20-30\,GeV$).  Alternatively, one might argue that there is a
stronger energy dependence related to some threshold behaviour for
putting the charm quarks on their  mass shell. We make a crude
model for this by taking a constant fraction  of the pQCD charm cross
section, where such a threshold factor is included,  to be our second
case of the intrinsic charm cross section  
\begin{equation}
{\rm IC2:}\;\;\; \sigma_{IC}(s) =  0.1\: \sigma_{pQCD}(s)
\label{eq:IC2}
\end{equation}
as shown by curve IC2 in Fig.\,\ref{fig:cross}. This is similar to the
low energy ($\sqrt{s}=20-40\,GeV$) treatment in~\cite{Vogt91}. The
normalisation is fixed to be the same as IC1 at the low energy
where evidence is claimed for intrinsic charm \cite{Vogt92}. There is,
however, some indication against such an increased cross section, as in
IC2, since no evidence for $J/\psi$ from intrinsic charm was found in
an  experiment \cite{noIC} at a somewhat higher energy ($800\,GeV$
proton beam energy).

In Fig.\,\ref{fig:cross} we have also compiled various data on the
charm  production cross section in proton-(anti)proton collisions at
various energies.  A few comments on the data in  Fig.\,\ref{fig:cross}
are here in order.  A given experiment is only sensitive to some
channels and a limited  kinematical region. The total charm cross
section is therefore obtained by a rescaling with charm decay branching
ratios and by using assumed  shapes of the $x_F$ distributions to
extrapolate to unmeasured regions.  In particular, corrections to
points 1,2,6 and 7 are small  while they are large for point 9 and 13. 
The bands 8,10,11 and 12 illustrate the uncertainties in these 
experiments due to this extrapolation.  In band 8 the uncertainty
includes a scaling for including $D^{\pm}$-mesons (taken from
\cite{lebc800, E789}). Data-band 11 is based on $D^+\bar{D}$
identification, 12 on $\Lambda_c^+\bar{D}$ and 10 on $\Lambda_c^+$ 
identification. Furthermore, points 3, 4 and 5 are from beam dump
experiments on heavy nuclear targets without direct charm
identification and have an additional  uncertainty from the scaling
with nuclear number. In point 4 the scaling $A^{0.75}$ has been
assumed, which we have rescaled in 4' to a $A^1$-dependence in order to
be consistent with the other beam dump  experiments and with our
model \cite{GIT}.  
Data points 2 and 6 come from $pp$ interactions with explicit
charm particle identification.   Although these issues leave some
uncertainty for each individual result,  the combination of all data
should give  a trustworthy knowledge on the charm cross section and its
energy dependence.

These inclusive charm cross section data can be reasonably well
understood by pQCD and does not leave much room for an IC component.
This follows from our detailed Monte Carlo study \cite{GIT} of pQCD
charm  production in high energy hadron collisions based on the {\sc
Pythia} program \cite{Jetset} and applied to high energy cosmic ray
interactions. The calculation is based on the
conventional folding of the parton densities in the  colliding hadrons
and the leading order QCD matrix elements, for the  gluon-gluon fusion
process $gg\to c\bar{c}$ and the quark-antiquark annihilation process
$q\bar{q}\to c\bar{c}$. Thus, the cross section is 
\begin{equation} \label{SIGMACC}
\sigma = \int \int \int dx_1dx_2d\hat{t} \;\; f_1(x_1,Q^2)\; f_2(x_2,Q^2)\;  
         \frac{d\hat{\sigma}}{d\hat{t}}
\end{equation} 
where $x_i$ are the parton longitudinal momentum fractions in the
hadrons  and the factorisation scale is taken as  $Q^2=(m^2_{\bot
c}\,+\,m^2_{\bot \overline{c}})/2$. The parton level pQCD cross section
$\hat{\sigma}$ depends on the Mandelstam  momentum transfer $\hat{t}$.
Next-to-leading-order corrections are known  and give a correction
which can be approximately taken into account by  an overall factor
$K=2$. The charm quark mass threshold  $\hat{s}=x_1x_2s > 4m_{c}^{2}$ 
is important and is fully included in  the matrix elements. The
dominating contribution to the cross section comes from the region 
close to this threshold, since $d\sigma /d\hat{s}$ is a steeply falling
distribution. It is therefore  important to  use QCD matrix elements
with the charm quark mass explicitly  included. The numerical value
used is $m_{c}=1.35\,GeV/c^2$ together with  $\Lambda_{QCD}=0.25\,GeV$
(from $MRS\,G$ \cite{mrsg}).

The results of this pQCD calculation is also shown in
Fig.\,\ref{fig:cross}.  At the highest energies, the parton densities
are probed  down to $x\sim 10^{-5}$ or even below.  The recent data
from  HERA \cite{hera1,hera2} show a  significant increase at small
$x$, $xf(x)\sim x^{-a}$ and constrain the  parton densities down to
$x\sim 10^{-4}$.  These data have been used in the  parameterisation
$MRS\,G$ \cite{mrsg} of parton densities resulting in the  small-$x$
behaviour given by the power $a=0.07$ for sea quarks and $a=0.30$  for
gluons. This is the most recent parameterisation, using essentially
all  relevant experimental data and can be taken as a standard choice. 
The effect on the total charm production cross section from the choice
of parton density parameterisation is illustrated in
Fig.\,\ref{fig:cross}. The result is shown with $MRS\,D_0$ \cite{mrsd0}
with small-$x$ behavior $x\,f(x)\sim const$, which before the HERA data
was an acceptable parameterisation. At high energy there is a large
dependence on the choice of parton density functions. The difference
between the $G$ and the $D_0$ parameterisations  should however not be
taken as a theoretical uncertainty. First of all the $D_0$
parameterisation is known to be significantly below the small-$x$ HERA
data and gives therefore a significant underestimate  at large
energies. Secondly, the naive extrapolation of the $G$ parameterisation
below the  measured region $x\gsim 10^{-4}$ at rather small $Q^2$
($\sim m_c^2$)  leads to an overestimate. A flatter dependence like
$x^{-\epsilon}$ with $\epsilon \simeq 0.08$ as $x\,\to\,0$ can be
motivated (\cite{SS} and references therein) based on a connection to
the high energy behaviour of cross sections in the Regge framework. The
implementation of this approach in {\sc Pythia} makes a smooth
transition  to this dependence such that the parton densities are
substantially  lowered for $x\lsim 10^{-4}$ leading to a substantial
reduction of the  charm cross section at large energies, as given by
the solid curve in Fig.\,\ref{fig:cross}. 

The pQCD calculation gives a quite decent agreement with experimental charm
production data over a wide range of energies. 
Nevertheless, the uncertainties in the data and the calculations cannot exclude 
some smaller non-perturbative contribution.
Charm production in pQCD is theoretically well defined and has only some
limited numerical uncertainty due to parameter values and NLO corrections 
which, however, can be examined and controlled. 
Non-perturbative contributions to charm production are, however, not 
theoretically well defined due to the
general problems of non-perturbative QCD. It is therefore, reasonable to take
pQCD as the main source of charm and consider, \eg, intrinsic charm as an 
additional contribution. 
We show this contribution in Fig.\,\ref{fig:cross} based on the two assumed 
energy dependences IC1 and IC2. Both cases can be consistent with the 
indications for IC mentioned in the Introduction, except the data in \cite{noIC}
which disfavours the stronger energy dependence of IC2.
One should note in this context, that an even stronger energy dependence would
be needed if the old ISR data were to be interpreted with intrinsic charm as
the dominating source.  

\subsection{Charm distributions at the Tevatron and LHC}

\begin{figure}[b]
\begin{center}
\includegraphics[width=16cm]{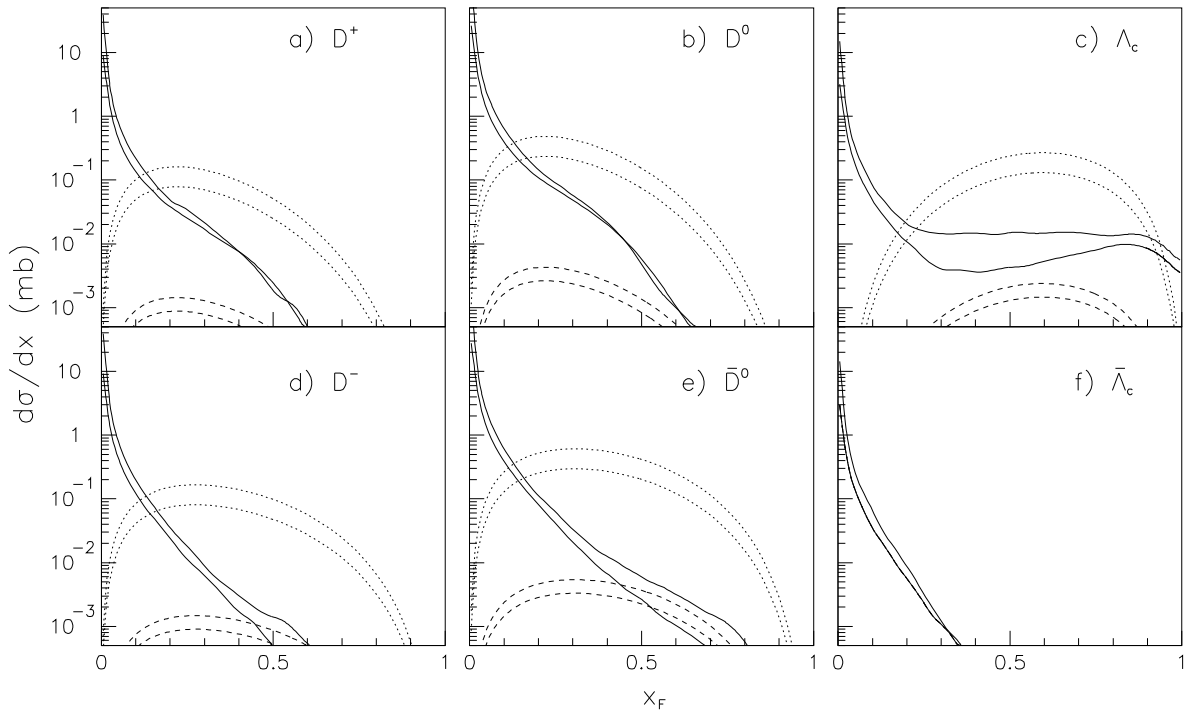}
\caption[junk]{\it 
Cross section versus longitudinal momentum fraction $x_F$ for charm
particles produced from pQCD (full curves) and from the intrinsic charm
model with energy dependence IC1 (dashed curves) and IC2 (dotted
curves). In each case, the upper curves correspond to LHC ($pp$ at
$\sqrt{s}=14\,TeV$) and the lower ones to the Tevatron ($p\bar{p}$ at
$1.8\,TeV$). Only one hemisphere $x_F>0$ (proton beam direction) is
given.}
\label{fig:tevatronx}
\end{center}
\end{figure}

Given the differences in the pQCD and the IC mechanisms, one expects 
characteristic differences in the spectra of produced charmed hadrons
at collider energies.  Charm produced through the pQCD mechanisms
should emerge with rather small longitudinal momentum or $x_F$. 
This results from the  parton fusion being largest close to
threshold $\hat{s}=sx_1x_2\sim4m_c^2$.  In contrast, intrinsic charm is
giving rise to charm particles at large  fractional momenta relative to
the beam particles, as explained before.  The latter process may
therefore dominate at some large $x_F$, with the  cross-over point
depending on the relative normalisations of the cross sections for the
two processes. 

\begin{figure}[t]
\begin{center}
\includegraphics[width=16cm]{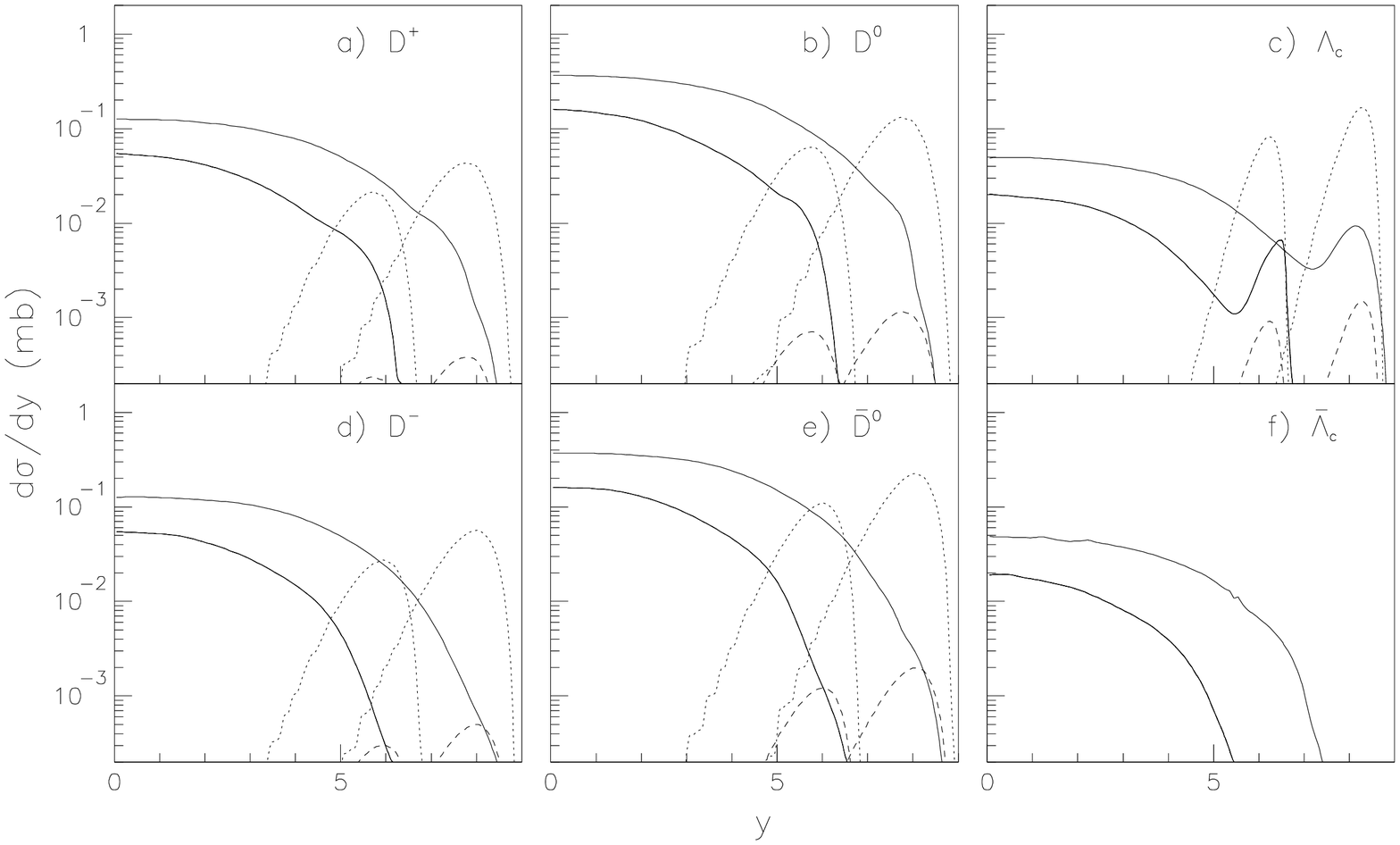}
\caption[junk]{\it 
Cross section versus rapidity for charm particles produced from pQCD 
(full curves) and from the intrinsic charm model with energy dependence
IC1 (dashed curves) and IC2 (dotted curves). In each case, the upper
curves correspond to LHC ($pp$ at $\sqrt{s}=14\,TeV$) and the lower
ones to the Tevatron ($p\bar{p}$ at $1.8\,TeV$). Only one hemisphere
$y>0$ (proton beam direction) is given.}
\label{fig:tevatrony}
\end{center}
\end{figure}

To study this, we have used our models and calculated the spectra in
$x_F$ and rapidity of charmed hadrons at the Tevatron and LHC shown in
Figs.~\ref{fig:tevatronx} and \ref{fig:tevatrony}. Only one hemisphere
($x_F>0, y>0$) is shown, corresponding to the proton beam direction at
the Tevatron, such that the other hemisphere is obtained with charge
conjugation symmertry. Since the pQCD processes are fully Monte Carlo
simulated one can easily extract the $x_F$ and rapidity
$y=ln\{(E+p_z)/(E-p_z)\}$. Our model for intrinsic charm is analytic
and formulated in terms of $x_F$-dependencies. The rapidity is then
calculated using $y\simeq ln\,(2x_F P/m_{\perp})$,  with $P$ the beam
momentum and $m_{\perp}$ the transverse mass of the charmed hadron. As
usual in IC models, we neglect the transverse momentum although it may
be expected to be of the same order as the charm quark mass. Including
$p_{\perp}$ fluctuations of this magnitude, would only cause shifts to
lower rapidity of about $ln(\sqrt{2})\simeq 0.35$ which is
significantly smaller than the widths of the rapidity peaks in
Fig.~\ref{fig:tevatrony} and therefore  not change the results
significantly. 

In comparison to the Tevatron, the charm cross section at LHC is larger
by a factor four for pQCD and IC2 and a factor two for IC1. The IC
distributions essentially scale in $x_F$, but shift to larger rapidity
at LHC due to the higher beam momentum. The increase for pQCD is mainly
at small $x_F$, due to the charm threshold moving to smaller momentum
fractions $x$ in the colliding particles. 

With the stronger energy dependence of IC2, the IC cross section is 
significantly  higher than that from pQCD at large longitudinal
momentum.  The milder energy dependence of IC1 gives a charm cross
section which is in general much smaller than the pQCD one and cannot
really compete even at high $x_F$.  The only exception is for
$\overline{D}$ where the previously discussed leading
particle effects are important, resulting in cross sections of similar 
magnitude as pQCD for very large $x_F$ or rapidity.  Thus, it is will
be very hard to test the IC1 case, but the IC2 case  could be
observable provided that the forward coverage of the detectors is
extended far enough. This could be considered in connection with
dedicated heavy flavour experiments covering the forward region in
particular. 

\section{Summary and conclusions}

The hypothesis of intrinsic charm quark-antiquark pairs in the proton 
wave function is not ruled out by experiment. On the contrary, there
are some evidence in favour of it, but no safe conclusion can be made
at  present. It is therefore important to consider various ways to
gain  additional information that could help clarify the situation.
Based on  previous work, we have constructed explicit models for
intrinsic charm  and how it may be examined in deep inelastic
lepton-nucleon scattering and  hadron collisions. Our models are partly
implemented in terms of Monte Carlo programs, which allow detailed
information to be extracted since complete events are simulated. 

In DIS such intrinsic charm quarks can be directly probed and we find
that it may dominate the inclusive charm $F_2^c$ structure function at
large $x$. Muons from charm decay can be used to signal events with
charm and we compare results of the IC mechanism with the conventional
perturbative QCD boson-gluon fusion process, as well as background
muons from $\pi$ and $K$ decays. We devise cuts in order to enhance IC
relative to the backgrounds. Signal and backgrounds can then be brought
to about the same level. We point out that data samples already collected by
NMC would be suitable for this purpose and could be sensitive down to
a level of about 0.2\% probability of intrinsic charm in the proton.  

Direct scattering on intrinsic charm quarks at HERA has been
investigated before \cite{IJN}. Here, we investigate the indirect
process with scattering  on a light valence quark such that the
$c\bar{c}$ fluctuation in the proton  remnant is realised. The
rapidities of the produced charmed particles have  been calculated and
found to be very forward, as expected. Present detectors do not have
enough forward coverage to detect these processes, but one may consider 
them in connection with possible upgrades for future HERA running.  

For hadronic interactions, the intrinsic charm model gives definite
predictions for charmed particle $x_F$ spectra. The absolute
normalisation and its energy dependence is, however, not clear. We have
investigated this in comparison with conventional pQCD productions
mechanism for charm and measured charm production cross sections. This
constrains the allowed energy variation of the IC cross section. Using
two simple models for this energy dependence, 
we calculate the $x_F$ and rapidity
distributions for charmed particles at the Tevatron and LHC. We find
that it is only if the IC cross section has a significant
rise with energy, that it can compete with normal pQCD production of
charm. In any case, the IC contribution is at very large forward
momenta, such that its detection would require coverage at very forward
rapidities.  

In the context of very high energy hadronic collisions one should also 
consider cosmic ray interactions in the atmosphere, which we have investigated
in detail \cite{GIT}. Here also, the intrinstic charm mechanism would 
contribute significantly only if it has a strong energy dependence.

Finally, one should remember the possible extension from intrinsic charm to 
intrinsic bottom quarks. Although we have not presented numerical results,
our methods can easily be applied for intrinsic bottom processes. 
In comparison to charm the general expectations are, as discussed in the 
Introduction, that the $x_F$-distributions will be only slightly harder 
but the overall rates lower by about a factor ten. 

\vspace{5mm}
\noindent
{\bf Acknowledgements:} We are grateful to S.~Brodsky for helpful 
communications and to J.~Rathsman for a critical reading of the manuscript.

\end{document}